  \documentclass[final,twocolumns,5p,times,sort&compress,12pt]{elsarticle}

\usepackage{graphicx}
\usepackage{dcolumn}
\usepackage{bm}
\usepackage{epstopdf}
\usepackage{bm,hyperref,color,breakurl}

	\usepackage{fancyheadings}
\renewcommand{\thispagestyle}[1]{} 

\begin{document}



\title{Magnetization distribution in a spin ladder-shaped quantum nanomagnet}


\author[a1]{K. Sza\l{}owski\corref{cor1}}
\ead{kszalowski@uni.lodz.pl}
\author[a1]{P. Kowalewska}
\address[a1]{Department of Solid State Physics, Faculty of Physics and Applied Informatics,\\
University of \L\'{o}d\'{z}, ulica Pomorska 149/153, 90-236 \L\'{o}d\'{z}, Poland}

\cortext[cor1]{Corresponding author}

\date{\today}

\begin{abstract}
The quantum nanomagnets show interesting site-dependent magnetic properties as a function of the temperature and the external magnetic field. In the paper we present the results of calculations for a finite quantum spin ladder with two legs, consisting of 12 spins $S=1/2$, with open ends. We describe our system with isotropic quantum Heisenberg model and perform exact numerical diagonalization of the Hamiltonian to use canonical ensemble approach. Our analysis focuses on the site-dependent magnetization in the system, presenting magnetization distributions for various interaction parameters. We discuss extensively the temperature and magnetic field dependences of individual site magnetizations. The interesting behaviour, with pronounced non-uniformity of magnetization across the ladder, is found. 
\end{abstract}

\maketitle

\section{Introduction}

The quantum magnetic nanosystems exhibit numerous non-trivial properties \cite{Himpsel1998,Owens2015}. Although the zero-dimensional nature of such systems excludes the presence of the typical phase transitions expected in infinite systems, yet a range of interesting phenomena specific to this class of objects can be observed instead. On the one hand, such nanomagnets can be experimentally realized either chemically, as molecular magnets \cite{Schnack2004,Friedman2010,Sieklucka2017}, or by assembling them on the surface atom by atom \cite{Khajetoorians2012}. On the other hand, their finite size enables the application of the most powerful (but computationally demanding) method for theoretical studies - the exact diagonalization, which yields an entirely physically correct picture, free from any artefacts even for fully quantum models \cite{Schnack2005,Weisse2008,Lauchli2011}. These facts should be supplemented with observation that nanomagnets can carry huge potential for applications in information storage and processing, both at classical level \cite{Khajetoorians2011,Loth2012,Yan2017} and at quantum level \cite{Leuenberger2001,Meier2003,Meier2003a,Affronte2007,Hoogdalem2014}. As a consequence, the studies of nanomagnetic systems are strongly motivated. This motivation seems to peak at the finite chain-like and ladder-shaped systems, which recently attract particular experimental attention \cite{Loth2012,Yan2017}. Although the theoretical studies of finite systems were mainly aimed at extrapolation to infinite, one-dimensional cases \cite{Cabra1997,Cabra1998,Honecker2000}, yet also chains and ladders of finite length are interesting by themselves (to mention, for example, the presence of nontrivial edge states) \cite{Machens2013,Lounis2008,Politi2009,Qin1995,Miyashita1993}.

It should be emphasized that the finite, zero-di-mensional nanomagnets exhibit lack of translational symmetry, so that all the physical quantities which are defined for single spins can be expected to be site-dependent. This contrasts with the behaviour of the infinite systems, where the symmetry of the magnetic ordering does not lead usually to such non-uniformity. Therefore, nanomagnets offer a particularly interesting opportunity to investigate the highly non-uniform systems. Especially, the magnetization can be expected to depend on the considered site, so that the study of the magnetization distribution is of primary importance. The local magnetization can be characterized experimentally with atomic-resolved \linebreak methods \cite{Hirjibehedin2006,Yayon2007,Meier2008,Wiesendanger2009,Serrate2010,Wiesendanger2011,Ternes2017} and experimental studies focused on the magnetization distribution \cite{Micotti2006,Ghirri2009,Guidi2015} can be mentioned. Moreover, the non-uniformity of magnetization can influence the functioning of any nanomagnet-based device or even serve as a basis for its design. The modeling of site-dependent magnetization as a function of the temperature and external magnetic field for cluster-like systems appears therefore well motivated and valuable, especially if based on the exact approach. We can mention that the theoretical studies of magnetization distribution (performed with either exact or approximate methods) are known in the literature both for zero-dimensional magnets \cite{Antkowiak2013} as well as for other non-uniform systems, like, for example, thin films \cite{Balcerzak1990,Wiatrowski1991,Zasada2007}. Also the thermodynamics of magnetic clusters was subject of several computational works exploiting the exact (or close to exact) approaches, involving both 'classical', Ising-based systems \cite{Richter1982,Zukovic2014,Zukovic2015,Zukovic2018,Karlova2017,Strecka2015} as well as highly non-trivial quantum Heisenberg systems \cite{Richter1982,Schmidt2005,Schnack2005,Schnack2005a,Konstantinidis2009,Schmidt2010,
Schnack2013,Schnack2013a,Ummethum2013,Langwald2013,Hanebaum2015,Schnack2016,
Irons2017}.

In view of the mentioned facts, the aim of our paper is to perform an exact study of the site-dependent magnetization of a selected ladder-shaped, finite nanomagnet. In particular, we would like to uncover the evolution of the magnetization distribution as a function of the temperature and the external magnetic field. The study is aimed at supplementing and developing the previous ground-state results \cite{Szalowski2018} obtained for a two-legged spin ladder composed of 12 spins. It should be emphasized here that our system of interest lacks translational symmetry because of the open ends of both chains (legs of the ladder). The effect of various intra- and interleg interactions of either ferro- or antiferromagnetic sign and different magnitudes will be characterized. Some further discussion concerning the selection of the system of interest can be found in the section \ref{Final remarks}.

\section{Theoretical model}

\begin{figure}[ht!]
  \begin{center}
   \includegraphics[width=0.95\columnwidth]{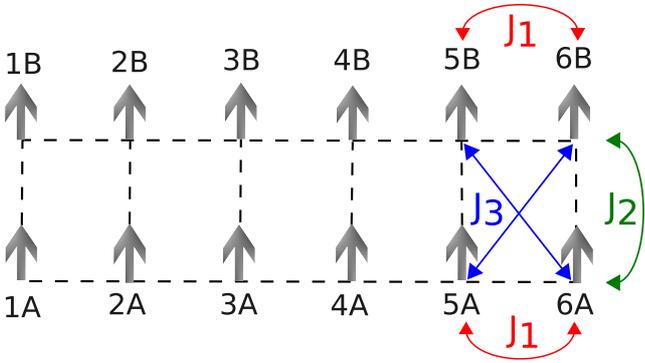}
  \end{center}
   \caption{\label{fig:1} A schematic view of the system of interest - a quantum nanomagnet being a finite two-legged ladder. The spins $S=1/2$ are labelled with the index of leg (A or B) and the position in the leg (1 to 6). The interactions between the spins are depicted schematically. }
\end{figure}

The study is devoted to the system being a spin ladder with two legs of finite length, consisting in total of $N=12$ quantum spins $S=1/2$ \cite{Szalowski2018}. Fig.~\ref{fig:1} presents a schematic view of the investigated nanomagnet. Each spin is labelled with an index of a leg (A or B) as well as the position in the leg ($i,j=1,...,6$). The interactions between the spins are isotropic in spin space and described with Heisenberg model, with the exchange integrals explained in Fig.~\ref{fig:1}. The system is ruled by the following Hamiltonian:

\begin{eqnarray}
\label{eq:1}
\boldsymbol{\mathcal{H}}&=&-J_1\left(\sum_{\left\langle iA,jA\right\rangle}^{}{\mathbf{S}_{iA}\cdot \mathbf{S}_{jA}}+\sum_{\left\langle iB,jB\right\rangle}^{}{\mathbf{S}_{iB}\cdot \mathbf{S}_{jB}}\right)\nonumber\\
&-&J_2\sum_{\left\langle iA,jB\right\rangle}^{}{\mathbf{S}_{iA}\cdot \mathbf{S}_{jB}}-J_3\sum_{\left\langle\left\langle iA,jB\right\rangle\right\rangle}^{}{\mathbf{S}_{iA}\cdot \mathbf{S}_{jB}}\nonumber\\&&-H\left(\sum_{iA}^{}{S^{z}_{iA}}+\sum_{iB}^{}{S^{z}_{iB}}\right). 
\end{eqnarray}

The intraleg coupling between nearest neighbours is denoted by $J_1$, whereas analogous interleg (rung) coupling amounts to $J_2$. In addition, interleg (crossing) interactions between second neighbours are denoted by $J_3$. The external magnetic field, defining the $z$ direction in spin space, is equal to $H$. The spin operators $\mathbf{S}=\left(\frac{1}{2}\boldsymbol{\sigma}^{x},\frac{1}{2}\boldsymbol{\sigma}^{y},\frac{1}{2}\boldsymbol{\sigma}^{z}\right)$ are composed of appropriate Pauli matrices.
 
The Hamiltonian (Eq.~\ref{eq:1}) and all the other quantum operators related to the system in question can be expressed as the matrices of the size 4096 $\times$ 4096. The exact numerical diagonalization of the Hamiltonian yields the eigenvalues $\epsilon_k$ and eigenvectors $\left|\psi_k\right\rangle$. On such basis, within the canonical ensemble approach, the statistical sum for the system in question can be expressed as 
\begin{equation}
\mathcal{Z}=\sum_{k}^{}{\exp\left(-\beta\, \epsilon_k\right)},
\end{equation}
where $\beta=\left(k_{\rm B}T\right)^{-1}$, with $k_{\rm B}$ denoting the Boltzmann constant. 
Also the thermal average value of an arbitrary quantum operator $\boldsymbol{A}$ can be directly determined from the formula:
\begin{equation}
\left\langle \boldsymbol{A}\right\rangle = \frac{1}{\mathcal{Z}}\sum_{k}^{}{\left\langle \psi_k\right| \boldsymbol{A}\left|\psi_k\right\rangle \exp\left(-\beta\, \epsilon_k\right)}.
\end{equation}
 
In the present paper the quantity of particular interest is the magnetization. The magnetization at $j$-th site can be expressed with the following operator:
\begin{equation}
\boldsymbol{m} ^{z}_j= \bigotimes_{i}^{}{ \left(\frac{1}{2}\delta_{i,j}\,\boldsymbol{\sigma}^z+\left(1-\delta_{i,j}\right)\boldsymbol{I}_2\right)}.
\end{equation}
The symbol $\otimes$ denotes the Kronecker (external) product, while $\delta_{i,j}$ is the Kronecker delta. The operator $\boldsymbol{\sigma}^{z}$ is the appropriate Pauli matrix and $\boldsymbol{I}_2$ is identity matrix of the size 2 $\times$ 2.
The total magnetization of the system is expressed as the following sum:
\begin{equation}
\boldsymbol{m} ^{z}_{T}= \sum_{j}^{}\boldsymbol{m} ^z_{j}.
\end{equation}
The key quantities studied in the present paper are: the average total magnetization $m^{z}_{T}=\left\langle \boldsymbol{m} ^{z}_{T} \right\rangle$ and the average magnetizations for individual sites of the nanomagnet, $m^{z}_{j}=\left\langle \boldsymbol{m} ^{z}_{j} \right\rangle$. The numerical results concerning their behaviour as a function of the temperature and external magnetic field will be extensively analysed in the next section of the paper.

\section{Numerical results and discussion}

\begin{figure}[ht!]
  \begin{center}
\includegraphics[width=0.95\columnwidth]{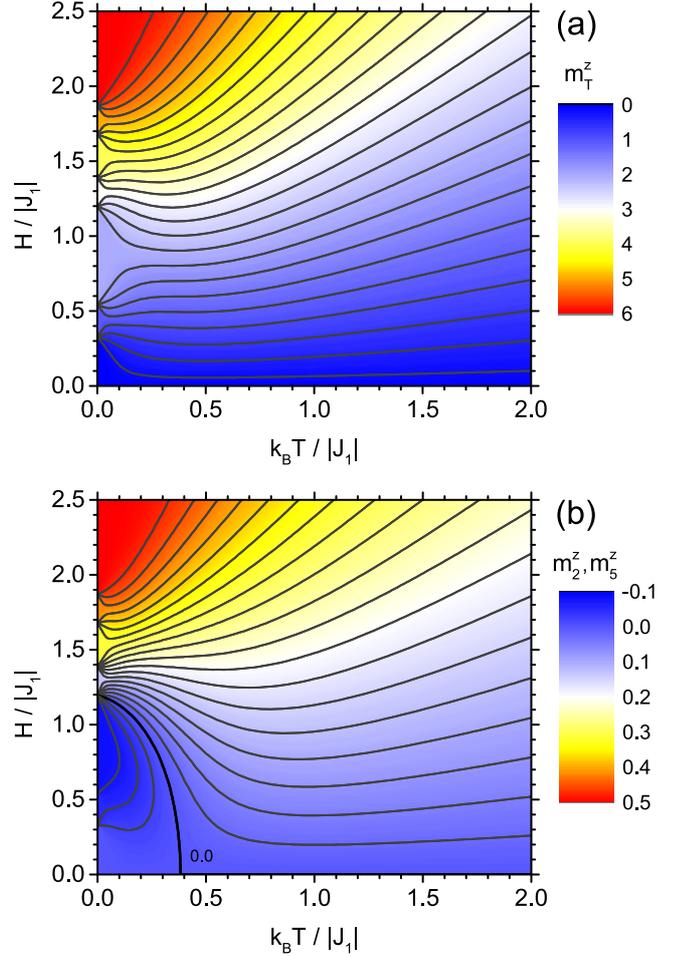}
  \end{center}
   \caption{\label{fig:2} The contour plots of total magnetization (a) and site-specific magnetizations for sites 2 and 5 (b) as a function of normalized temperature and normalized magnetic field, for the interaction parameters $J_1<0$, $J_2/|J_1|=0.5$ and $J_3/|J_1|=0$. For labelling of the sites see Fig.~\ref{fig:1}.}
\end{figure}

\begin{figure}[ht!]
  \begin{center}
\includegraphics[width=0.95\columnwidth]{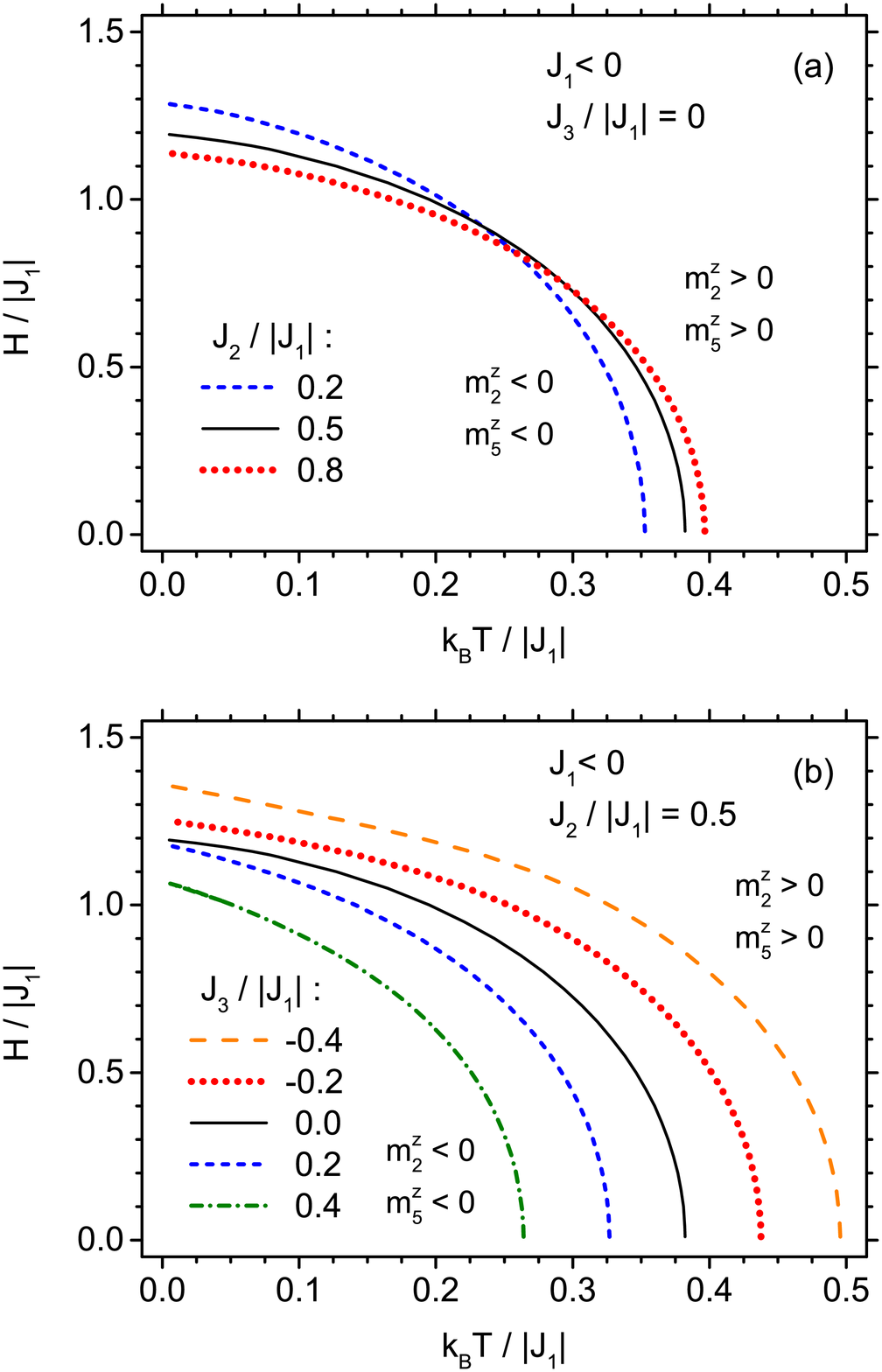}
  \end{center}
   \caption{\label{fig:new} The temperature-magnetic field phase diagram showing the boundaries between the ranges where $m^z_2,m^z_5<0$ and where $m^z_2,m^z_5>0$, for $J_1<0$. The case (a) corresponds to $J_3/|J_1|=0.0$ and various values of $J_2/|J_1|$; the case (b) corresponds to $J_2/|J_1|=0.5$ and various values of $J_3/|J_1|$. }
\end{figure}

\begin{figure*}[ht!]
  \begin{center}
\includegraphics[width=1.9\columnwidth]{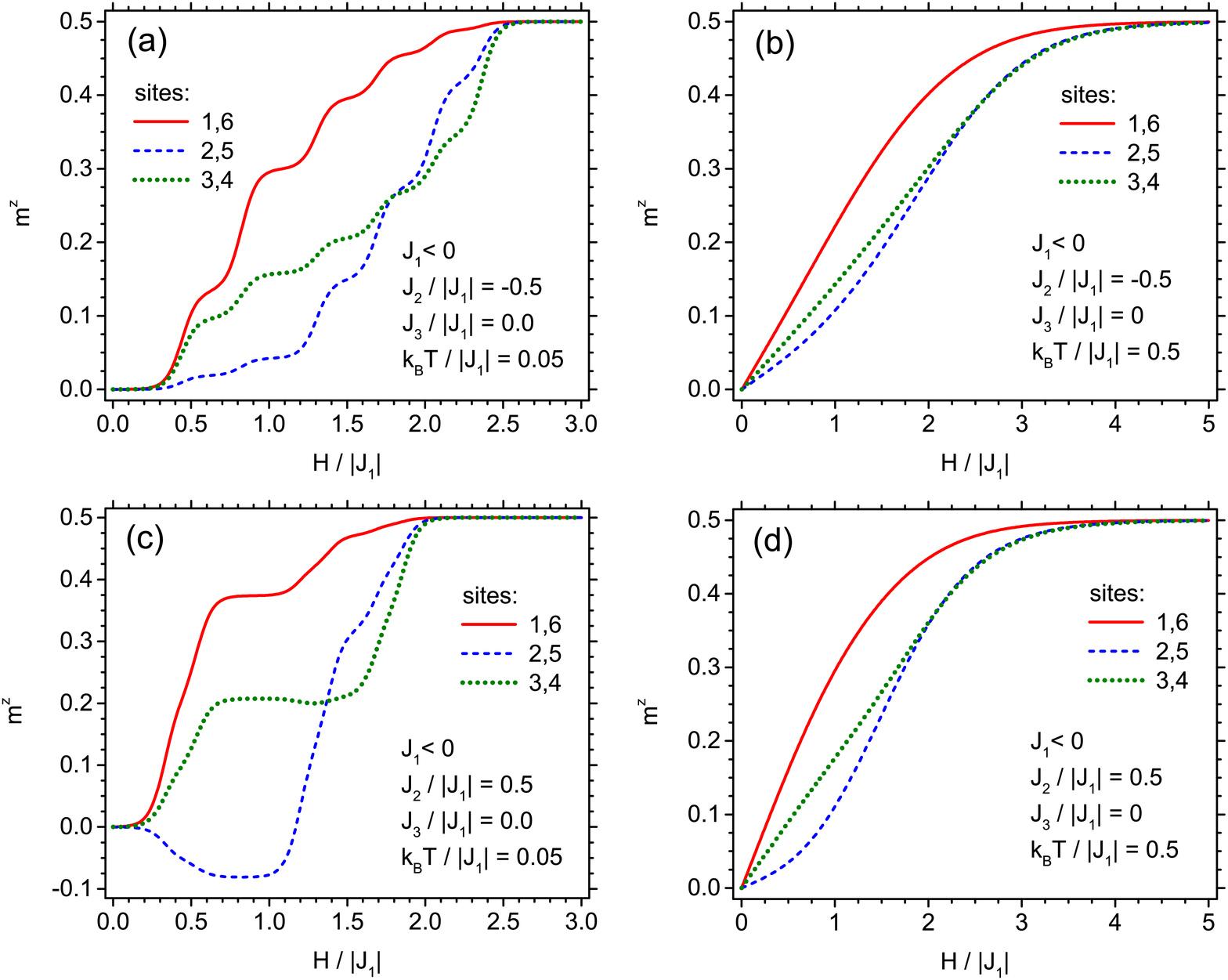}
  \end{center}
   \caption{\label{fig:3} The site-specific magnetization (for all inequivalent sites - see Fig.~\ref{fig:1}) as a function of the normalized magnetic field for low ($k_{\rm B}T/|J_1|=0.05$) and high ($k_{\rm B}T/|J_1|=0.5$) normalized temperature, for $J_1<0$ and for either $J_2/|J_1|=-0.5$ ((a), (b)) or $J_2/|J_1|=0.5$ ((c), (d)).}
\end{figure*}

\begin{figure}[ht!]
  \begin{center}
   \includegraphics[width=0.9\columnwidth]{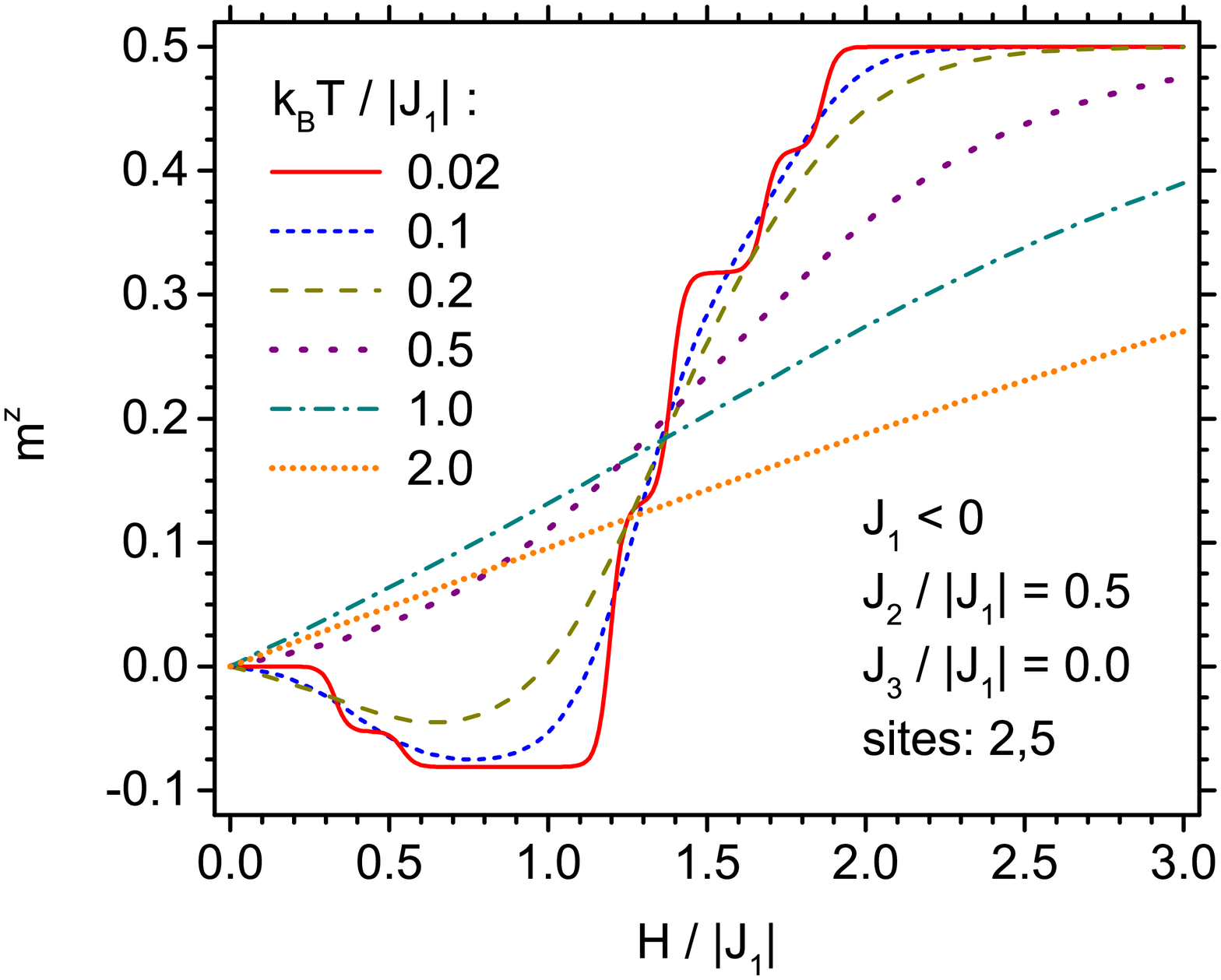}
  \end{center}
   \caption{\label{fig:4} The site-specific magnetization for the inner sites of the ladder (2,5) (see Fig.~\ref{fig:1}) as a function of normalized magnetic field, for a wide range of normalized temperatures, for $J_1<0$ and $J_2/|J_1|=0.5$.}
\end{figure}

All the results of numerical calculations presented in this section were obtained using Wolfram Mathematica software \cite{Wolfram2010}. The discussed diagrams concern in general the dependence of magnetization on the temperature and the magnetic field for the system in question, for a representative selection of the interaction parameters between the spins. For the investigated range of parameters, no site-dependent magnetization was found to depend on the leg index (A or B) and only the dependence on the position in the leg was observed.

Let us commence the analysis from the dependence of the total magnetization of the system on the temperature and magnetic field, which is shown in Fig.~\ref{fig:2}(a) for the case of $J_1<0$ and $J_2>0$. The density plot allows additionally to trace the contours of constant magnetization vs. both thermodynamic variables - $T$ and $H$. The values lay between 0 and 6, where the value of 0 is achieved for $H=0$ and arbitrary temperature and 6 means the magnetic saturation. The points in which numerous contours tend to merge at the ground state ($T=0$) correspond to the subsequent critical magnetic fields at which the total magnetization changes its value discontinuously. Such behaviour is presented in Fig.~2(a) in our previous work (Ref.~\cite{Szalowski2018}) and the values of critical magnetic field shown there are consistent with the limiting behaviour seen in Fig~\ref{fig:2}(a) in the present work. At finite temperatures, it is evident that the total magnetization always increases with an increasing magnetic field. However, analysis of the isolines of constant magnetization supports the statement that in some range the magnetization increases with the increasing temperature and then falls down, reaching some local maximum. In some narrow ranges also a minimum and maximum or two maxima separated by a minimum or a plateau can exist (as evidenced by the detailed analysis of the data). 

After a brief analysis of the total magnetization, let us focus the attention on the particular position in the ladder. Namely, in Fig.~\ref{fig:2}(b) we present the temperature and magnetic field dependence of the magnetization for the sites in ladder labelled with 2 and 5 (see Fig.~\ref{fig:1} for explanation). In such case it is seen that the magnetization can reach both positive values and negative values (in some area of the diagram for low temperatures and magnetic fields, as limited with a bold contour). It follows that the magnetization for temperatures low enough drops down when the magnetic field increases and then crosses the zero value and increases further. Therefore, a non-monotonous behaviour is predicted at sufficiently low temperatures. In addition to extrema achieved as a function of the magnetic field for constant temperature also extrema as a function of the temperature for fixed magnetic field can be expected. 

The occurrence of antiparallel orientation of spins at the sites 2 and 5 with respect to the other spins, promoted by external magnetic field, is a phenomenon which deserves a somehow more detailed analysis. Therefore, Fig.~\ref{fig:new} was prepared to illustrate its sensitivity to the interaction parameters $J_2$ and $J_3$. In Fig.~\ref{fig:new}(a) the boundary $m^z_2=m^z_5=0$, separating the ranges where $m^z_2,m^z_5<0$ and where $m^z_2,m^z_5>0$ is plotted as a function of the temperature and magnetic field, for $J_3/|J_1|=0.0$ and for varying $J_2/|J_1|$. Please note that the line for $J_2/|J_1|=0.5$ corresponds to the analogous contour of $m^z_2=m^z_5=0$ plotted in bold in Fig.~\ref{fig:2}(b). It is visible that increase in rung ferromagnetic coupling $J_2$ reduces the magnetic field range at low temperatures where $m^z_2,m^z_5<0$ whereas the corresponding temperature range at low magnetic fields is expanded. The effect of the crossing interaction $J_3$ on the analogous phase diagram is illustrated in Fig.~\ref{fig:new}(b) for $J_2/|J_1|=0.5$. In this case it is evident that the ferromagnetic crossing couplings tend to reduce the range where $m^z_2,m^z_5<0$ (limiting both the temperature and the magnetic field), whereas antiferromagnetic $J_3$ interactions act in the opposite direction, causing the range to expand in temperature and magnetic field.

The discussion of the behaviour of magnetization in contour plots shown in Fig.~\ref{fig:2} motivates the interest in detailed analysis of its cross-sections, which will be presented in further plots.

Firstly let us analyse the behaviour of the site-dependent magnetization as a function of the normalized magnetic field. Such data are shown in Fig.~\ref{fig:3} for antiferromagnetic $J_1$. The cases Fig.~\ref{fig:3}(a) and (c) are for low temperature, whereas Fig.~\ref{fig:3}(b) and (d) present higher temperature results. 
Diagrams (a) and (b) are prepared for antiferromagnetic $J_2$. The magnetizations increase with the magnetic field. At low temperature, traces of a step-wise increase of magnetization are visible, which resemble the discontinuous magnetization steps separating the plateaux at the ground state. The differences between the magnetization values at different sites are rather pronounced and the dominant contribution originates from the sites at the ladder ends. The elevated temperature (b) causes the magnetization increase to become more regular and smooth (with conserved tendency of taking the largest values at the ladder ends).  
In the next pair of diagrams - Fig.~\ref{fig:3}(c) and (d) we have ferromagnetic $J_2$. Here the low-temperature behaviour of magnetization is very different from the one visible in Fig.~\ref{fig:3}(a) for all-antiferromagnetic couplings. Namely, the magnetizations at sites 2 and 5 take negative values as the magnetic field increases, then cross the zero value and increase quite rapidly to reach saturation. On the contrary, at the remaining sites the magnetizations are always positive - at the outermost sites they increase monotonically with the field, whereas for sites 3, 4 a narrow range of temperatures where magnetization decreases is noticeable. At intermediate fields a sort of long plateau emerges. For higher temperatures, all the magnetizations increase regularly and monotonically with the field - so that we observe fully analogous situation to that illustrated in Fig.~\ref{fig:3}(b) for $J_2<0$. Therefore, it can be deduced that along with the increase of temperature the behaviour of magnetization vs. magnetic field always becomes regular. Also for low temperatures, higher magnetic field must be applied to the system to reach the common saturation value of magnetization for all sites of the ladder. At lower temperatures the saturation is reached simultaneously for all the sites, whereas at higher temperature the ends of the ladder tend to saturate first.

The non-monotonical dependence of magnetization at several sites on the magnetic field, with change of the magnetization sign, is an interesting behaviour and it deserves a separate illustration. It is shown in Fig.~\ref{fig:4}, presenting the magnetization at sites 2 and 5 vs. the magnetic field for a wide range of temperatures. At the lowest studied temperature we observe traces of 6 step-wise changes (characteristic of ground state when the total spin changes discontinuously). These steps are smoothed by the thermal fluctuations. It is interesting that for several lower temperatures the magnetization takes negative values when the field increases and then crosses zero at some nonzero field. The range of fields below which a negative magnetization occurs is narrowed by the temperature; above some temperature the temperature dependence of magnetization becomes linearized and the magnetization itself is always positive. Also the saturation is reached for lower magnetic field at lower temperatures.

\begin{figure*}[ht!]
  \begin{center}
   \includegraphics[width=1.9\columnwidth]{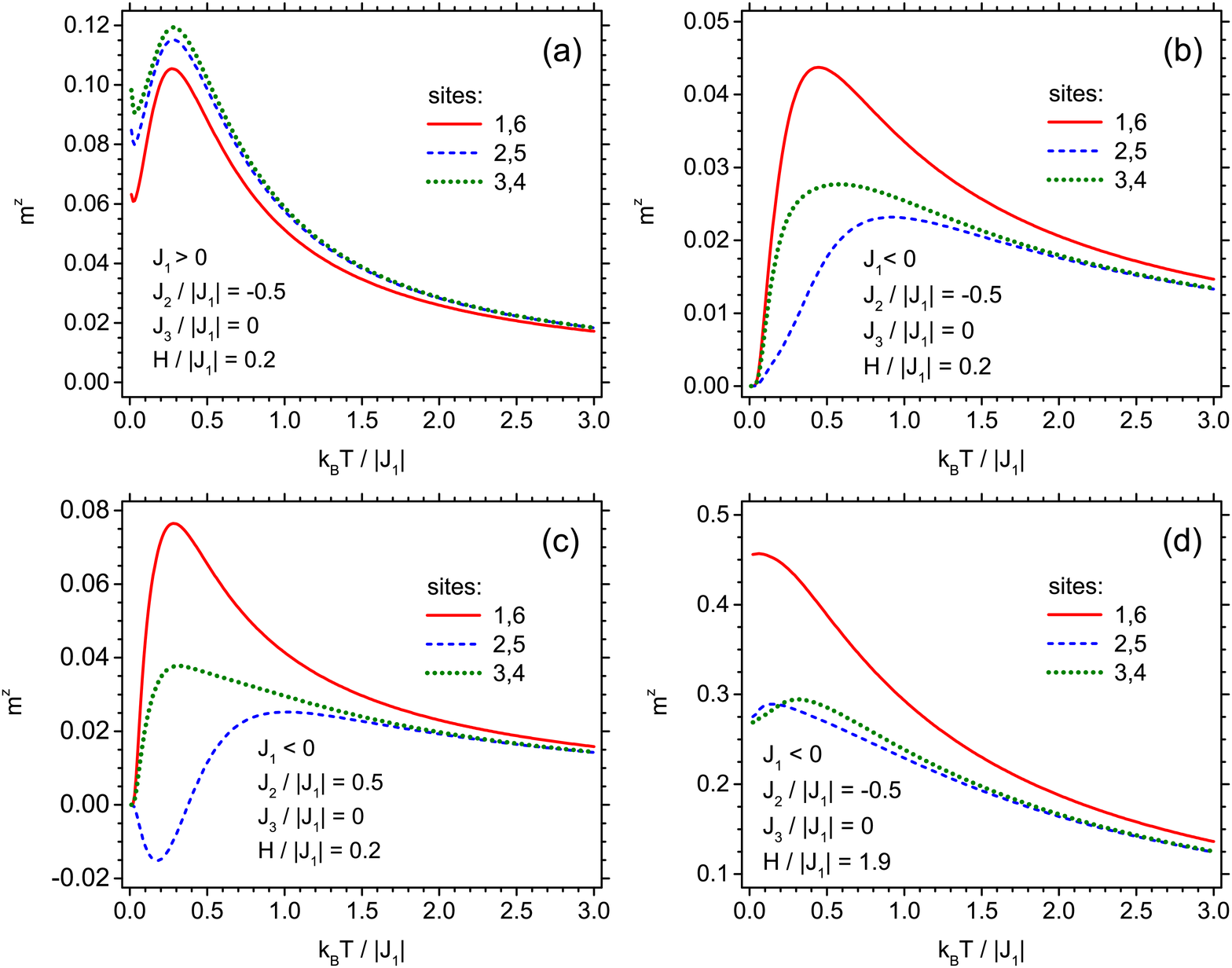}
     \end{center}
   \caption{\label{fig:5} The site-specific magnetization (for all inequivalent sites - see Fig.~\ref{fig:1}) as a function of the normalized temperature for low ($H/|J_1|=0.2$) and high ($H/|J_1|=1.9$) normalized magnetic field. The cases (a)-(d) correspond to various values of $J_1$ and $J_2$, as indicated in the panels.}
\end{figure*}

\begin{figure}[ht!]
  \begin{center}
   \includegraphics[width=0.95\columnwidth]{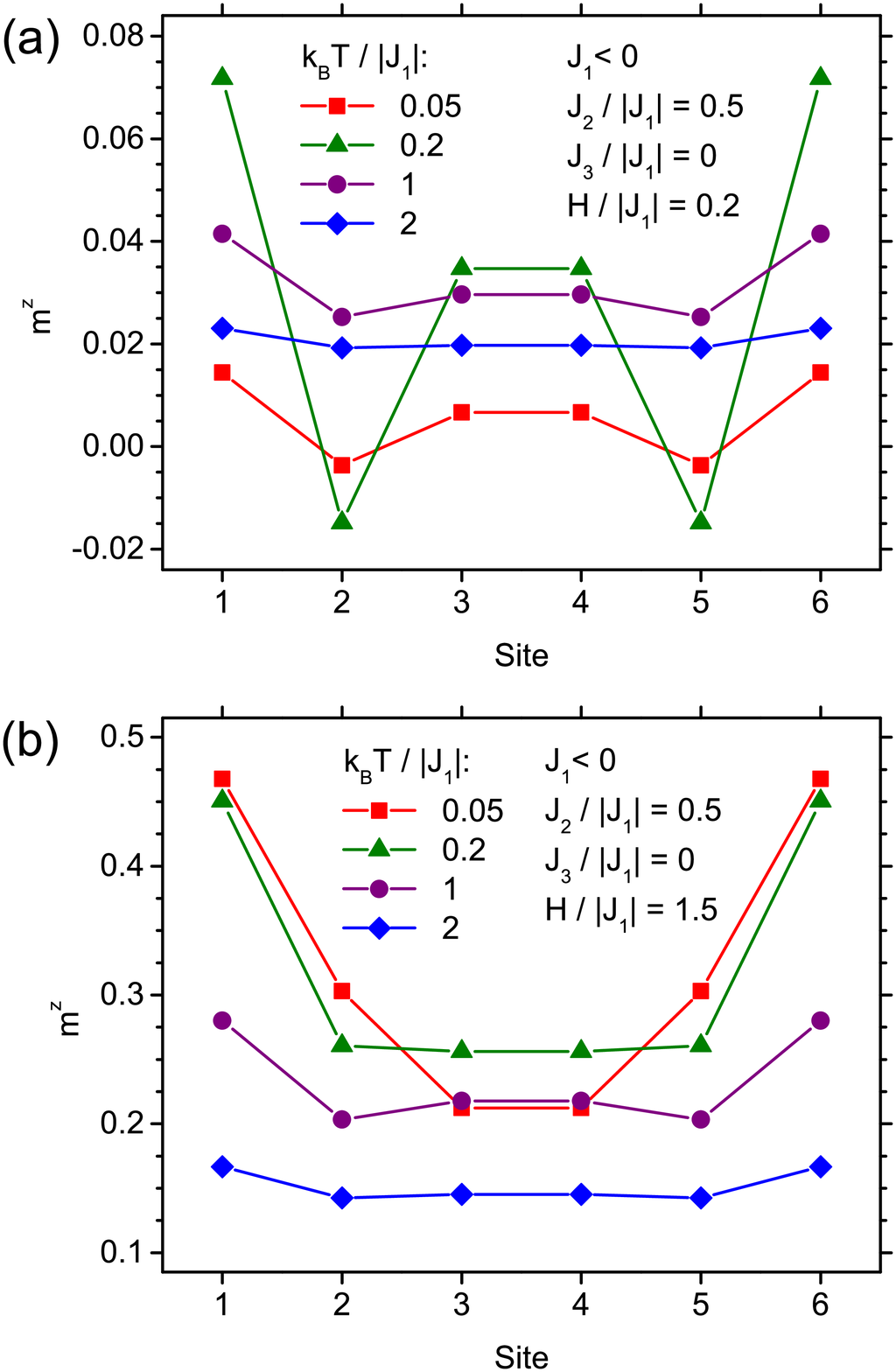}   
  \end{center}
   \caption{\label{fig:6} Magnetization distribution across the ladder (for numeration of sites see Fig.~\ref{fig:1}) for low ($H/|J_1|=0.2$) (a) and high ($H/|J_1|=1.5$) (b) normalized magnetic field, for a few normalized temperatures, for $J_1<0$ and $J_2/|J_1|=0.5$. }
\end{figure}

\begin{figure}[ht!]
  \begin{center}
   \includegraphics[width=0.95\columnwidth]{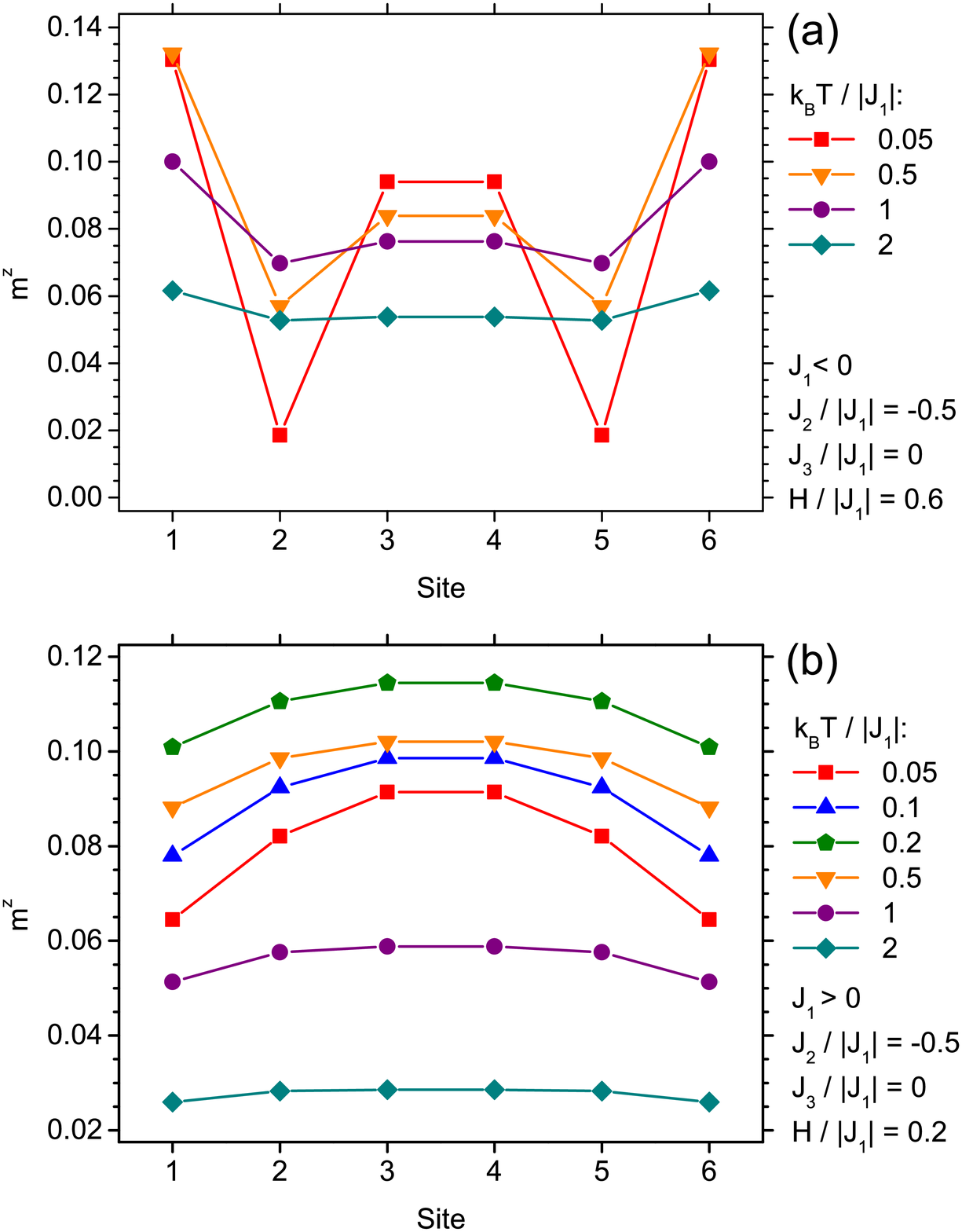}   
  \end{center}
   \caption{\label{fig:7} Magnetization distribution across the ladder (for numeration of sites see Fig.~\ref{fig:1}) for fixed normalized magnetic fields, for a few normalized temperatures, for $J_2/|J_1|=-0.5$ with $J_1<0$ (a) or $J_1>0$ (b).}
\end{figure}

\begin{figure}[ht!]
  \begin{center}
   \includegraphics[width=0.95\columnwidth]{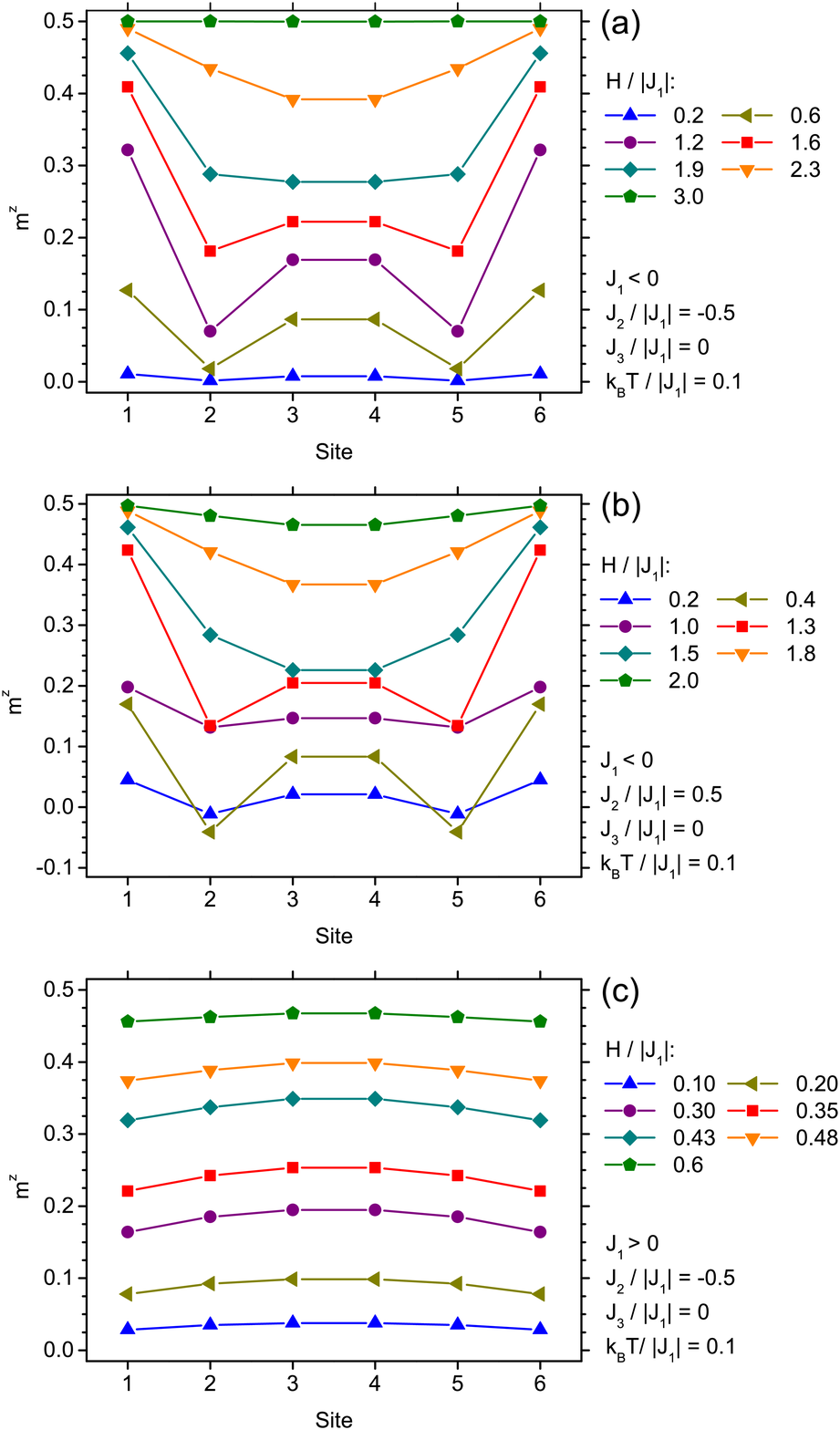}  
   \end{center}
   \caption{\label{fig:8}  Magnetization distribution across the ladder (for numeration of sites see Fig.~\ref{fig:1}) for fixed low normalized temperature $k_{\rm B}T/|J_1|=0.1$ and a few normalized magnetic fields. The cases (a)-(c) correspond to various values of $J_1$ and $J_2$, as indicated in the panels. }
\end{figure}

The temperature dependences of magnetization for specified sites can be traced in Fig.~\ref{fig:5} for some representative selection of interaction parameters $J_1$ and $J_2$ and for various values of magnetic field (weak or strong one). First two plots - Fig.~\ref{fig:5}(a) and (b) - allow the comparison of the cases of $J_1<0$ and $J_1>0$ at $J_2/|J_1|=-0.5$ for the same, weak magnetic field. For the ferromagnetic case the behaviour of magnetization is quite complicated - it first drops (starting from the nonzero value, since at the ground state the spin at this magnetic field is equal to 1). Then it rises, reaching a broad maximum and finally drops to low values at high temperatures. It should be observed that the largest magnetization is always in the middle of the ladder, while it is weakest at the ends (but the differences are not greatly pronounced). For $J_1<0$ the magnetization first rises (from zero value) with increasing temperature and then drops (significantly slower than for the ferromagnetic case). Contrary to the previous case, the values of magnetization peak at the ends of the ladder, while at the intermediate sites 2 and 5 they reach minima. Moreover, the differences between various sites are very significant (and overall the magnetization values are lower). The situation is even more striking for $J_1<0$ and ferromagnetic $J_2$, as illustrated in Fig.~\ref{fig:5}(c). There, the magnetization for sites 2 and 5 develops a minimum with negative values when the temperature rises, whereas at the remaining sites a maximum with positive values is noticed. Then, the magnetization at intermediate sites crosses zero and develops a broad maximum; for higher temperatures the magnetizations at all sites tend gradually to zero. Finally, in Fig.~\ref{fig:5}(d) the case of all antiferromagetic couplings is shown for high magnetic field. There, a weak maximum of magnetization is observable at low temperatures and then a decrease is seen when the temperature increases. The magnetization is strongest for the exterior sites of the ladder, while its ratio to the remaining magnetizations depends on the temperature (but the differences for the sites inside the ladder are not well pronounced). 

The last type of graphs presents the magnetization profiles along each ladder leg. It should be emphasized that the magnetization is identical for the corresponding sites in both legs of a quantum ladder, denoted by A and B in Fig.~\ref{fig:1}. It can be generally stated that every distribution of magnetization is symmetric with respect to the symmetry axis of the ladder. This feature is present due to the fact that we consider ladder legs (chains) with open ends. Let us note that the lines connecting the symbols are guides to eyes only. 

In Fig.~\ref{fig:6} the distribution of magnetization values across the ladder can be followed for various temperatures, for the case of antiferromagnetic $J_1$ and ferromagnetic $J_2$ ($J_2/|J_1|=0.5$). The two plots are compared to emphasize the differences in magnetization for low and high external magnetic field. For low field (Fig.~\ref{fig:6}(a)), which at the ground state would correspond to the total spin equal to 0, an interesting behaviour can be noticed, when for sites 2 and 5 the magnetization is negative for lower temperatures (as already seen in Fig.~\ref{fig:3} and Fig.~\ref{fig:4}(c)). This reflects the antiferromagnetic correlations between spins at sites 2 or 5 and the neighbouring spins in the same ladder leg. Due to the even number of sites in each leg, two sites closest to the symmetry axis remain ferromagnetically correlated. Therefore, the profile has two deep minima which tend to become increasingly shallow and eventually flatten when the temperature increases. The magnetization behaves non-monotonically as a function of the temperature (as seen in Fig.~\ref{fig:4}(c)). For strong field (which corresponds to the total spin of 4 at the ground state) the profile is slightly different. Namely, it exhibits in general a single, pronounced minimum in the middle of the ladder (so mainly the sites at ladder ends magnetize). Again, the magnetization first rises and then drops when the temperature increases; the profile tends also to flatten. Of course, the magnetizations have much bigger values in Fig.~\ref{fig:6}(b) than in Fig.~\ref{fig:6}(a).   

In the next plot, Fig.~\ref{fig:7} we have contrasted the cases of $J_1<0$ and $J_1>0$ (for the same $J_2/|J_1|=-0.5$) to notice crucial differences between these cases. The magnetization profiles are again plotted for various temperatures and, this time, for such magnetic fields that at the ground state the total spin is equal to 1. 

For $J_1<0$ (Fig.~\ref{fig:7}(a)) at low temperatures a pair of minima exists, like in the case of Fig.~\ref{fig:6}(a). When the temperature rises the profile tends to take a more convex shape. In general, the highest magnetization is observed at the ends of the ladder.

On the other hand, for the case of $J_1>0$ (Fig.~\ref{fig:7}(b)) the magnetization distribution is always concave, with maximum values in the middle of the ladder. The magnetization at the ends is significantly reduced for lower temperatures and then the profile gradually flattens. The magnetization is a non-monotonic function of the temperature, as it is seen in Fig.~\ref{fig:5}(a) - it first increases and then falls down rapidly.

Finally, let us analyse the charts which juxtapose the magnetization profiles at constant low temperature $k_{\rm B}T/|J_1|=0.1$ at various magnetic fields (their subsequent values would correspond at the ground state to all the possible values of the total spin, i.e. from 0 to 6 - see Fig.~1 in Ref.~\cite{Szalowski2018}). Fig.~\ref{fig:8} compares both signs of $J_1$ and $J_2$ (with an exception of all ferromagnetic couplings). In almost all the cases, all the magnetizations increase with the increase of the external magnetic field. The behaviour of magnetization at sites 2 and 5 at low fields for $J_1<0$ and $J_2/|J_1|=0.5$ (Fig.~\ref{fig:8}(b)) is at variance with the usual tendency (already demonstrated, for example, in Fig.~\ref{fig:4}). The general shape of the magnetization profiles for $J_2<0$ and $J_2>0$ for antiferromagnetic $J_1$ is similar; however, for purely antiferromagnetic couplings the behaviour of magnetization is more regular and all the sites have the same orientation of magnetizations. For lower fields the profiles exhibit two symmetric minima, and when the field increases they merge and the magnetization distribution becomes convex, with a central, pronounced minimum. On the contrary, for $J_1>0$, as seen in Fig.~\ref{fig:8}(c), the profile is always concave, with a single maximum. Moreover, the magnetizations are much more uniform across the ladder than in the cases of $J_1<0$.  

\section{Final remarks}
\label{Final remarks}
In the paper the quantum ladder consisting of 12 spins was analyzed. The system of interest had open ends, what strongly promotes the inhomogeneity in the distribution of the local properties, due to lack of translational symmetry. Summarizing the key results we can indicate that the investigated quantum nanomagnet exhibits interesting site-dependent magnetic properties. We modelled our ladder-shaped system with an isotropic quantum Heisenberg model with various intra- and interleg interactions. All the results were obtained with exact numerical diagonalization technique within canonical ensemble formalism. The external magnetic field was included. The calculations were performed for finite temperatures, to supplement our previous study of the same system focused solely on the ground state \cite{Szalowski2018}. The cases of ferromagnetic and antiferromagnetic couplings were compared (considering all the exchange integrals $J_1$, $J_2$ and $J_3$). We focused on the range of $J_2$ and $J_3$ where a dominant coupling is $J_1$, i.e. when the system can be considered as a pair of coupled finite chains (ladder legs), not a system of coupled dimers (ladder rungs). We observed a non-uniform magnetization distribution depending on the ladder site (symmetric with respect to the ladder symmetry axis due to the open ends) and followed its evolution when varying either the temperature or the field. In particular, we found a non-monotonic variability of magnetizations as a function of the the temperature with pronounced extrema for some interaction parameters. Moreover, we also uncovered the antiferromagnetic ordering induced by the magnetic field for some range of couplings (with opposite sign of interleg and intraleg interaction). All the results show that a rather rich behaviour of magnetization distribution can be expected for such magnetic nanosystem. Let us mention that the experimental site-resolved magnetization studies as a function of the magnetic field have been reported for a finite chain \cite{Micotti2006,Guidi2015}, proving the usefulness of theoretical modelling of local magnetization distribution in nanomagnets.

Finally, let us comment on the selection of our system of interest, which is composed of spins $S=1/2$. Frequently the molecular nanomagnets carry larger spins at metallic sites, however, numerous examples of systems with spin equal to 1/2 can also be given \cite{Furrer2013}. Moreover, the nanomagnetic systems assembled via STM-based techniques can be composed of atoms exhibiting spin equal to 1/2, as already demonstrated experimentally \cite{Yang2017,Otte2008}. In addition, there exist examples of compounds for which a spin $S=1/2$ ladder with two legs is a commonly accepted model, to mention compounds containing Cu ions \cite{Azuma1994,Watson2001} or V ions \cite{Isobe1998}. It might be expected that nanostructures based on such compounds might also constitute ensembles of zero-dimensional ladder-shaped magnets. Other examples of clusters composed of spins-1/2 can be found in Ref.~\cite{Haraldsen2005}. Therefore, the interest in zero-dimensional systems composed of spins-1/2 is justified from the experimental/design point of view.

Focusing the attention on the behaviour of the systems with spin-1/2 constituents can be also motivated by the fact that most clear quantum effects can be expected for the case of lowest spin whilst, in general, larger spins tend to behave in more classical manner. Moreover, spin-1/2 systems are described by the Hamiltonians with relatively low number of free parameters, what facilitates a more systematic study across the interaction parameters space.

Further research may concern other thermodynamic properties of this system at finite temperatures in the presence of the field. Also it is desirable to study zero-dimensional systems with different geometry (for example to capture even-odd effects) or (much more computationally demanding) systems with higher spins.

\section*{Acknowledgments}

\noindent This work has been supported by Polish Ministry of Science and Higher Education on a special purpose grant to fund the research and development activities and tasks associated with them, serving the development of young scientists and doctoral students.


\bibliographystyle{elsarticle-num}


\end{document}